\documentstyle[preprint,aps,a4,fullpage,subeqn,epsf,here,rotate]{revtex}

\newcommand{\fur}{\qquad\mbox{for }\,}

\newcommand{\beq}[1]{\begin{equation}\label{#1}}
\newcommand{\eeq}{\end{equation}}
\newcommand{\beqa}[1]{\begin{eqnarray}\label{#1}}
\newcommand{\eeqa}{\end{eqnarray}}
\newcommand{\bsq}[1]{\begin{subequations}\label{#1}}
\newcommand{\esq}{\end{subequations}}

\newcommand{\gl}[1]{eqn.\ (\ref{#1})}
\newcommand{\gls}[2]{eqs.\ (\ref{#1}) and (\ref{#2})}
\newcommand{\glto}[2]{eqs.\ (\ref{#1}) to (\ref{#2})}
\def\gtrapprox{\raise2.5pt\hbox{$>$}\llap{\lower2.5pt\hbox{$\approx$}}}
\def\lssapprox{\raise2.5pt\hbox{$<$}\llap{\lower2.5pt\hbox{$\approx$}}}

\begin{document} 

\title{Equations of structural relaxation}

\author{M. Fuchs and Th. Voigtmann}
\address{
Physik-Department, Technische Universit\"at M\"unchen, 
D-85747 Garching, Germany} 
\date{\today}
\maketitle

\begin{abstract} 
In the mode coupling
theory of the liquid to glass transition the long time 
structural relaxation follows from equations solely determined by equilibrium
structural parameters. The present extension of these structural
relaxation equations to arbitrarily short times on the one
hand allows calculations unaffected by model assumptions about the microscopic 
dynamics and on the other hand supplies new starting points for analytical 
studies. As a first application,
power--law like structural relaxation at a glass--transition singularity
is explicitly proven for a special schematic MCT model. 
\end{abstract} 
\vfill
\leftline{Contribution to the proceedings of the conference:}
\leftline{{\it 
Seventh international Workshop on Disordered Systems}, Molveno, March
1999}
\bigskip\bigskip
\rightline{Submitted to Philosophical Magazine}
\newpage

\centerline{{\Large {\bf \S 1. Introduction}}}
\bigskip

Dense liquids exhibit anomalously slow, temperature sensitive and
non-exponentional relaxation processes. They are
conventionally termed `structural relaxation'
because cooperative rearrangements of particles are presumably involved.
The mode coupling theory (MCT) suggests that the evolution
 of the anomalous
dynamics and its splitting off from the normal liquid dynamics is caused
by bifurcation singularities in non--linear retarded equations of motion for
the (normalized) intermediate scattering functions $\Phi_q(t)=S_q(t)/S_q$
(Leutheusser 1984, Bengtzelius et al. 1984).
These functions are some of the most simple ones specifying structural
dynamics.  The bifurcations of MCT, and the long--time dynamics in their
vicinity, are solely determined by the equilibrium (static) structure factor
$S_q$, see (G\"otze and Sj\"ogren 1989, Franosch et al. 1998), and thus the MCT
provides a frame to single out and to define structural relaxation. In this
contribution the MCT equations for the structural relaxation are extended to
arbitrarily short times thus rendering them well defined even in the limit that
the microscopic transient dynamics can be neglected. Whereas in (Franosch et
al. 1998) this was achieved by a discrete--dynamics model, here integral
equations with explicit initial conditions are formulated. Thus the dependence
on the initial conditions is displayed clearly. Additionally, monotone solutions
are guaranteed (Fuchs et al. 1991). Furthermore, for the obtained structural
dynamics
equations non--exponential relaxation  can be proven for a special MCT model.

\bigskip\bigskip
\centerline{{\Large {\bf \S 2. The equations of motion of  MCT}}}
\bigskip
 
The MCT equations of motion for the autocorrelation functions of density
fluctuations in simple liquids, $\Phi_q(t)$, are derived from Newton's
equations with the Zwanzig--Mori formalism leading to:
\beqa{e1}
& \partial_t^2 \Phi_q(t) + \Omega_q^2 \Phi_q(t) + \int_0^t dt'\;  M_q(t-t')\;
\partial_{t'} \Phi_q(t') = 0 \; , &  \\ 
& \Phi_q(t\to 0) = 1 - \frac 12 (\Omega_q t)^2 \label{e2} \; , &
\eeqa
and subsequent mode coupling approximations for the generalized longitudinal
viscosity finally resulting in:
\beqa{e3}
& M_q(t) = \Omega^2_q m_q(t) + M^{\rm reg.}_q(t) \; , & \\
& m_q(t) = \sum_{{\bf k}+{\bf p}={\bf q}}\; V({\bf q};{\bf k},{\bf p})\,
\Phi_k(t) \Phi_p(t) & \; . \label{e4}
\eeqa
In \gl{e1}, the frequency $\Omega_q$ sets the microscopic time
scale for the bare oscillators with wave vector
$q$. In \gl{e3}, viscous damping arises from mode--mode coupling, $m_q(t)$,
\gl{e4}, and from microscopic short--time dynamics denoted by
$M^{\rm reg.}$; see (G\"otze 1991) for explicit expressions and discussions.
 A Laplace transformation of
\gls{e1}{e3} using $\hat{\Phi}(s)=\int_0^\infty dt\,  e^{-st}\,
\Phi(t)$ leads to
\beq{e5}
\hat{\Phi}(s) =   1\, /\, [\; s + 1\, /\, [\; \hat{m}(s) + (s+\hat{M}^{\rm
reg.}(s))/\Omega_q^2\; ]\,] \; .
\eeq
The existence of a unique solution of
the MCT equations  which has all properties of an autocorrelation function has
been  shown by Haussmann (1990). Further and more detailed properties could be
proven by G\"otze and Sj\"ogren (1995) upon the assumption of overdamped short
time motion. Let us mention two results relevant for this work,
 namely $(i)$ the existence
of bifurcations in the equations derived from \gls{e1}{e4} for the long--time
limits, $f_q=\lim_{t\to\infty} \Phi_q(t)$; here the $f_q$ are the glass form
factors and the bifurcations happen for  critical values of the mode
coupling vertices, $V=V^c$, where $f_q=f^c_q$. $(ii)$ for regular vertices, the
$f_q$ depend smoothly on the $V$, and there exists a longest relaxation time,
$1/\Gamma$, cutting off the final relaxation exponentially, $\Phi_q(t) - f_q =
{\cal  O}(e^{-\Gamma t})$ for $V\ne V^c$.  

One of the most simple models exhibiting the generic
features of the MCT bifurcation dynamics in \glto{e1}{e4} is obtained by the
simplification to study only one correlator, $\Phi(t)$, which experiences
non--linear feedback via one mode--coupling functional $m(t) = v_1 \Phi(t) +
v_2 \Phi^2(t)$ with $v_{1,2}\ge0$ (G\"otze 1984).  
We will start by discussing this so--called
$F_{12}$--model before generalizing our results to the equations
relevant for simple liquids.

\bigskip\bigskip
\centerline{{\Large {\bf \S 3. Results for the  schematic $F_{12}$--model}}}
\bigskip

`Equations of the structural relaxation' in the $F_{12}$--model are
equations of motion determined solely by the vertices $v_1$ and $v_2$ which 
play the role of the static structural information entering the MCT
vertices $V$ in \gl{e4}.  Bifurcations in the model lie on two lines
parametrized by the exponent parameter $\lambda$ 
 (G\"otze 1984): type--B lines,
where $f^c=1-\lambda>0$, at $v^c_1=(2\lambda-1)/\lambda^2$ and
$v_2^c=1/\lambda^2$ with $1/2\le\lambda<1$, and 
type--A transitions, where $f^c=0$, at $v^c_1=1$ and $v_2^c=\lambda$ for
$0\le\lambda\le1$. 
Whereas in \glto{e1}{e4} the
microscopic transient is required in order for solutions to exist 
(G\"otze 1991), after
partial integration the short--time variation can be neglected,
$\partial_t^2 \Phi(t)  + \int_0^t dt'\;  M^{\rm reg.}(t-t')\; \partial_{t'}
\Phi(t') \ll \Omega^2 \Phi(t)$, leading to:
\beq{e6}
\Phi(t) = m(t) - \frac{d}{dt} \int_0^t \; dt'\; m(t-t')\; \Phi(t')\; ,
\eeq
while the expression for the memory function is not changed:
\beq{e6b}
m(t) = v_1 \Phi(t) + v_2 \Phi^2(t)\; .
\eeq
Demanding the solutions to be of regular variation and that the Laplace
transforms $\hat\Phi(s)$ and $\hat m(s)$ exist, enabling one to use Tauber
and Abel theorems, see e.~g.\ (Feller 1971), the equations of structural
relaxation have to be completed by specifying the initial variation:
\beq{e7}
\Phi(t) \to (t/t_*)^{-1/3} \fur t \to 0\; .
\eeq
In this case, the convolution integral, which will be
abbreviated as $\int_0^tdt'm(t-t')\Phi(t')=(m*\Phi)(t)$, leads 
to $(m*\Phi)(t\to0)\to v_2 \,t_*\, B(1/3,2/3)$, where $B(x,y)$ is Euler's
Beta function. The Laplace transform of \gl{e6} for all $s$ then is:
\beq{e9}
(\, 1 + s\, \hat{m}(s)\, )\;  \hat{\Phi}(s) =  \hat{m}(s) + v_2 t_*
B(1/3,2/3)\;,
\eeq
It is a straightforward exercise to verify that the last term in \gl{e9}
is required for the existence of solutions with the requested properties
(Voigtmann 1998), thus leading to \gl{e7}.

%
%
Note, that the bifurcations to non--ergodic solutions cause
$\hat\Phi$ and $\hat m$ to increase for small frequencies, suggesting to
neglect $(s+\hat{M}^{\rm reg. })\Omega^{-2}\ll \hat{m}(s)$ in \gl{e5} in
order to derive equations for the structural relaxation alone. However,
this procedure misses the non--analytic behaviour of $(m*\Phi)(t\to0)$
connected to the limit $s\to\infty$.

One can extend \gl{e7} by a short--time series expansion, see appendix A,
\beq{e8}
\Phi(t) = (t/t_*)^{-1/3}\; (\,  1 + \sum_{n=1}\, c_n\; (t/t_*)^{n/3}\,  ) 
\fur t< r \, t_* \; . 
\eeq
Figure 1 shows the numerically obtained radii of convergence, $r$, 
of \gl{e8} along the line of bifurcations in the $F_{12}$--model.

Figure 2 shows numeric solutions of \glto{e6}{e7} which
exhibit slow relaxation processes upon approach of a bifurcation
singularity at critical values of the coupling vertices. This explains the
sensitive dependence of the dynamics on small regular changes in $\varepsilon$,
which measures the distance to the singularity. The algorithm described in
(Fuchs et al. 1991) is well suited for the numerical integration. A type--B
fold bifurcation at $\varepsilon=0$  separates ergodic, $\Phi(t\to\infty)=f=0$
for $\varepsilon<0$, from non--ergodic dynamics, $f\ge f^c>0$
for $\varepsilon\ge0$.  Asymptotic expansions identify two divergent time
scales connected with two power law relaxation windows (G\"otze 1984).
As  demonstrated e.~g.\ in (Franosch et al. 1998) 
the long--time solutions of the
complete MCT equations of motion, \glto{e1}{e4}, become independent of the
microscopic transient. For long times they agree with the solutions of the
equations of structural relaxation, \gls{e6}{e6b}. This is exemplified
in figures 2 and 3 using two approximations for the
regular microscopics in the $F_{12}$--model: undamped oscillation, $M^{\rm
reg.}\equiv0$, and pure relaxation, $(s+\hat{M}^{\rm
reg.}(s))\Omega^{-2}\to \Gamma_0$. It is apparent that oscillatory short--time
dynamics masks the structural relaxation more strongly than relaxational.

Assuming regular variation of the solutions of \glto{e1}{e4}, 
at a bifurcation point, power law relaxation could be argued:
\beq{e10}
\Phi(t) = f^c + h\; (t/t_0)^{-a}\; ,\fur t\to\infty\quad \mbox{and }\; 
v_{1,2}=v_{1,2}^c\; ,
\eeq
where the exponent $a$ follows from the exponent parameter
$\lambda$, which is a function of $v_{1,2}^c$,
 via $\lambda=B(1-a,1-a)(1-2a)$ (G\"otze 1984).  
From the equations of structural relaxation, \gls{e6}{e6b}, and the short--time
expansion \gl{e8}, one concludes that this asymptote requires  $t> t_* r$ in
general. The transient time $t_0$  is obtained from matching \gl{e10} to
\gl{e8}; while the critical amplitude $h=(1-f^c)$ for the $F_{12}$--model.

In the $F_{12}$--model there exist two points where the short--time
expansion can be found explicitly and solves the equations of structural
relaxation, \glto{e6}{e7}, for all times, i.e. $r\to\infty$ in 
\gl{e8}; see appendix A and figure 1: 
\beq{e11}
\Phi(t) =  f^c + (t/t_*)^{-1/3}\;, \fur
\lambda=\lambda_*={\textstyle \frac 13} B({\textstyle \frac 23},{\textstyle
\frac 23})=0.684\ldots \quad\mbox{and }\; 
v_{1,2}=v_{1,2}^c\; .
\eeq
Both points are bifurcation points, as finite distances 
from the singularities lead to the existence of long--time relaxation rates as
shown in appendix B, see also (G\"otze 1984, G\"otze 1991). Furthermore, 
the initial condition, \gl{e7}, requires for the critical exponent
$a=a_*=1/3$ which is satisfied for $\lambda=\lambda_*$. 
Thus, \gl{e11} verifies power--law relaxation at a bifurcation
singularity of equations describing the structural relaxation of a schematic
MCT model. 

The existence of a solution of the integral equations for all times at a
singular point also allows for these special cases to deduce a long--time
solution for small distances from the bifurcation in an expansion which can be
expected to have a finite radius of convergence, $(t/t_0) |\varepsilon|^{3/2} <
r_\beta$.  
\beq{e12}
\Phi(t) \to (t/t_0)^{-1/3}\, [\; 1 + \sum_{n=1} \; A_n \; (\, \varepsilon\,
(t/t_0)^{2/3}\, )^n\; ] \; , \fur t\to\infty \;.
\eeq
The calculation sketched in appendix B, thus introduces the
$\beta$--time scale, $t_\varepsilon=t_0 |\varepsilon|^{-1/2a}$ as $a=1/3$ for
$\lambda=\lambda_*$, and 
agrees with the expansion of the so--called $\beta$--correlator for small
rescaled times $\tilde{t}=t/t_\varepsilon$; see  (G\"otze 1984). There it is
also shown for general $\lambda$, how to extend this solution to even longer
times. Note that the expansion carried through in appendix B does not require a
matching  to the critical decay.

\bigskip\bigskip
\centerline{{\Large {\bf \S 4. Extensions to MCT equations for simple
liquids}}} 
\bigskip

The extension of the analysis of the bifurcation dynamics of schematic
models to the full MCT equations for simple liquids, \glto{e1}{e4}, has been
achieved in (G\"otze 1985) and rests upon the central--manifold concept. The
generalization of the equations of structural relaxation to coupled, wave
vector dependent equations is trivial; \gl{e6} attains $q$--indices and
\gl{e6b} is replaced by \gl{e4}. For quadratic polynomials in \gl{e4}, the
power law of the initial  condition is not  changed:
\beq{e13}
\Phi_q(t) \to x_q \; t^{-1/3} \; \fur t\to0\; ;
\qquad x_q\ge0\; .
\eeq
It is surprising that a vector $x_q$ has to be specified in order to
render the solutions unique; recall that owing to the factorization property
of the  critical decay, $\Phi_q(t) = f^c_q + h_q (t/t_0)^{-a}$ (G\"otze 1985), 
the long--time dynamics of the structural relaxation becomes unique by
specifying only the one matching time  $t_0$. 

From simple generalizations of the calculations presented in appendix A it is
obvious that the short--time expansion, \gl{e8}, can be obtained in the wave
vector dependent case also.

\bigskip\bigskip
\centerline{{\Large {\bf \S 5. Conclusions}}}
\bigskip

Definitions of equations for structural relaxation not restricted to long times
are given within the MCT. These equations, \gls{e4}{e6}, together
with the initial conditions, \gl{e13}, determine smooth but sensitively varying
intermediate scattering functions $\Phi_q(t)$ from the equilibrium
structure of the liquid; the static structure factor $S_q$ is the only quantity
appearing in the equations of motion.  The possibility to specify an arbitrary
initial non--negative amplitude, $x_q$, is worth stressing and also holds
for the discrete dynamics model of (Franosch et al. 1998). The short time
divergence of the solutions of the structural relaxation equations renders them
unphysical for too short times; they violate the imposed normalization of the
$\Phi_q(t)$. This stresses that the physical mechanism (the so--called
`cage effect') causing structural relaxation is relevant for  long--time
dynamics only. Any complete description of the density fluctuations  
requires a transient to regular microscopic short--time dynamics.
Nevertheless, extending the structural relaxation equations to all times on the
one hand allows the definition of a lower limit of validity of this long time
description --- the time where the solutions increase above unity could be
chosen --- and on the other hand presents a conceptual simplification as some
transient effects are eliminated. Thus for example, analytical proofs of
power--law relaxation at bifurcation points become possible for special cases;
recall that at regular points of the full MCT equations a longest
relaxation time exists. Also, the smaller of the two divergent time scales of
MCT could be identified from a perturbation expansion. Moreover, the
aspect that MCT equations of motion with different models for the short--time
microscopic motion lead to the identical long--time dynamics is again observed.

\bigskip\bigskip
\centerline{\Large \bf Acknowledgements}

We thank Prof. W. G\"otze for discussions and critical reading of the
manuscript. The work was supported by the Deutsche Forschungsgemeinschaft
under contract Fu 309/2-1 and by Verbundprojekt BMBF 03G05TUM.

\bigskip\bigskip
\centerline{{\Large {\bf Appendix A}}}
\bigskip
\newcounter{apgl}
\setcounter{apgl}{1}
\renewcommand{\theequation}{A\theapgl}

Entering the ansatz $\Phi(t)=f^c+\phi(t)$ into \gls{e6}{e6b} 
leads to 
\beq{a1}
\check{\sigma}=\bar{\sigma}\;  \phi(t) - \bar\lambda \;  \phi^2(t)
+(1/\lambda'+\tilde{\sigma})\; \dot{(\phi*\phi)}(t)+ \dot{(\phi^2*\phi)}(t)\; 
\eeq
\addtocounter{apgl}{1}
where a dot denotes a time--derivative and the following abbreviations are
used:  $\bar\lambda=\lambda/\lambda'$, $\lambda'=\lambda/(1-f^c)$,
$\check\sigma=(1-f^c)^2\sigma/\Delta$,
$\sigma=(1-f^c)(\delta v_1f^c+\delta v_2{f^c}^2)$,
$\Delta=(\lambda'+(1-f^c)^2\delta v_2)$,
$1/\lambda'+\tilde\sigma=[1+(1-f^c)^2(\delta v_1+2\delta v_2f^c)]/\Delta$,
and $\bar\sigma=(1-f^c)(1+\tilde\sigma-(1+\sigma)/\Delta)$
which are obtained from $v_{1,2}=v^c_{1,2}+\delta v_{1,2}$, where the
limit of $v_2$ approaching zero has been excluded.
In the vicinity of a type--B transition, $\lambda'=1$, $\bar\lambda=\lambda$.
Close to a type--A transition $\lambda'=\lambda$, and thus $\bar\lambda=1$.
From the short--time
expansion, \gl{e8}, one obtains a recursion relation for
$\tilde{c}_n$, where $\tilde{c}_0=1$, $\tilde{c}_1=c_1-f^c$ and
$\tilde{c}_n=c_n$ else, and $\delta_{n,m}$ is a Kronecker symbol. 
\beqa{a2}
\tilde{c}_{n+1} & = & \frac{3/(n+1)}{(2B({\textstyle
\frac{n+2}{3}},{\textstyle \frac 23})+B({\textstyle
\frac{n+3}{3}},{\textstyle \frac 13}))}\; \{ \; 
   \sum_{m=0}^n [ \bar\lambda- {\textstyle \frac{n+1}{3}} B({\textstyle
\frac{n-m+2}{3}},{\textstyle 
\frac{m+2}{3}})(1/\lambda'+\tilde{\sigma})]\,  \tilde{c}_{n-m}
\tilde{c}_m + \check\sigma\delta_{n,2} - \nonumber\\
& &
(1-\delta_{n,0}) [\, \bar{\sigma} \, \tilde{c}_{n-1} +
 {\textstyle\frac{n+1}{3}} \sum_{m=1}^n ( \sum_{k=0}^m B({\textstyle
\frac{n-k+2}{3}},{\textstyle \frac{k+2}{3}})\;
\tilde{c}_{m-k}\tilde{c}_k + B({\textstyle\frac{n+2}{3}},{\textstyle
  \frac{2}{3}})\tilde{c}_m )
\tilde{c}_{n+1-m} \, ]\; \}
\eeqa
\addtocounter{apgl}{1}
At two singular points, where
$\check{\sigma}=\bar{\sigma}=\tilde{\sigma}=0$  and $\lambda=\lambda_*=
B({\textstyle \frac 23},{\textstyle \frac 23})/3=0.684\ldots$, all
$\tilde{c}_{n}$ vanish.  This proves \gl{e11}. Approaching the  $F_1$--model
at $(v_1,v_2)=(v_1,0)$ (G\"otze 1991), the radius of convergence of \gl{e8}
shrinks as for a $m$--th order polynomial mode coupling functional a similar
expansion with  $\Phi(t)\to(t/t_*)^{-1/(m+1)}$ for $t\to0$ applies (Voigtmann
1998). 

\bigskip\bigskip
\centerline{{\Large {\bf Appendix B}}}
\bigskip
\setcounter{apgl}{1}
\renewcommand{\theequation}{B\theapgl}

Close to the type--B transition at $\lambda=\lambda_*$, 
the \gl{a1}  can be expanded around the solution
$\phi_0(t)=t^{-1/3}$ at $\delta v_{1,2}=0$. 
Using $\phi(t)=\phi_0(t)+\phi_1(t)$, specifying the path such
that $\tilde{\sigma}=0$ for simplicity and integrating  this leads to:
\beq{b1}
\int_0^t ds\, K(t,s) \, \phi_1(s) =  \check{\sigma} t - {\textstyle\frac
23}\bar{\sigma}  t^{2/3} -
((\bar{\sigma}+\phi_1+2\phi_0\phi_1)*\phi_1)(t) +
((\lambda_*-\phi_0-\phi_1)*\phi_1^2)(t)\; ,
\eeq
\addtocounter{apgl}{1}
where the Volterra kernel is of generalized Abel form:
\beq{b2}
K(t,s) =  2 \; (t-s)^{-1/3} - 2 \lambda_* \; s^{-1/3} + (t-s)^{-2/3} + 2\; 
(t-s)^{-1/3}\;  s^{-1/3}\; .
\eeq 
\addtocounter{apgl}{1}
From the short--time expansion \gl{e8}, $\phi_1(t\to0)\propto t^{1/3}$ can be
deduced in this case, ruling out that Kernel $K$ has  eigenvalue zero. 
The expansion $\phi_1(t)=\sum_{n=1}\, \sigma^n\; g^{(n)}(t)$ thus leads to a
recursive set of solvable  linear integral equations for the $g^{(n)}$.
If their long--time behaviour is searched for, only the first two terms in $K$
need to be considered and by induction one shows that $g^{(n)}(t)\to A_n\;
t^{(2n-1)/3}$ for $t\to\infty$, where the $A_n$ obey: $A_1=1/(2(B({\textstyle
\frac 23},{\textstyle \frac 43})-\lambda_*))$ and 
\beq{b3}
A_n = - \frac 12 [\, \sum_{m=1}^{n-1}\, A_{n-m} \, A_m \; (\, 
B({\textstyle \frac{2(n-m+1)}{3}},{\textstyle \frac{2m+2}{3}})-\lambda_*\;
{\textstyle \frac{2n+1}{3}} \,) \, ]/(
B({\textstyle \frac{2}{3}},{\textstyle \frac{2n+2}{3}})-\lambda_*\;
{\textstyle \frac{2n+1}{3}} \,) \; .
\eeq 
Replacing $\sigma$ with $\varepsilon$, this coincides with \gl{e12} and the
short--time expansion of the $\beta$--correlator for $\lambda=\lambda_*$
(G\"otze 1984).  

\newpage
\centerline{\large \bf References}
\bigskip

\noindent Bengtzelius, U., G\"otze, W., and Sj\"olander, A., 1994,
J. Phys. C., {\bf 17}, 5915.
\newline
\noindent Feller, W., 1971, {\it An Introduction to Probability Theory and its
Applications}, Vol. 2, (New\\ \hspace*{1cm} York: Wiley).
\newline
\noindent Franosch, T., G\"otze, W., Mayr, M. R.,  and Singh, A. P.,
1998,  J. Non--Cryst. Solids, \\ \hspace*{1cm} {\bf 235-237}, 66.
\newline
\noindent Fuchs, M., G\"otze, W., Hofacker, I.,  and Latz, A.,
1991,  J. Phys.: Condens. Matter, {\bf 3}, 5047. 
\newline
\noindent  G\"otze, W., 1984, Z. Phys. B, {\bf 56}, 139.
\newline
\noindent  G\"otze, W., 1985, Z. Phys. B, {\bf 60}, 195.
\newline
\noindent  G\"otze, W., 1991, {\it Liquids Freezing and Glass
Transition},   edited by  J. P. Hansen, D. Levesque\\ \hspace*{1cm} 
and J. Zinn-Justin (Amsterdam: North Holland), p. 287.  
\newline
\noindent  G\"otze, W., and Sj\"ogren, L.,  1989,   J. Phys.: Condens. Matter,
{\bf 1}, 4183.  
\newline
\noindent  G\"otze, W., and Sj\"ogren, L.,  1995, J. Math. Anal. Appl., {\bf
195}, 230. 
\newline
\noindent  Haussmann, R., 1990, Z. Phys. B, {\bf 79}, 143.
\newline
\noindent  Leutheusser, E., 1984, Phys. Rev. A, {\bf 29}, 2765.
\newline
\noindent  Voigtmann, Th., 1998, Diploma Thesis, Technische Universit\"at
M\"unchen, Garching.
\newpage
\bigskip\bigskip
\centerline{{\Large {\bf Figure captions}}}
\begin{itemize}
\item[Figure 1:] Radius of convergence $r$ of \gl{e8} numerically obtained
using the Cauchy--Hadamard criterion and \gl{a2} along the line of bifurcations
in the $F_{12}$--model. The line parametrized by $\lambda$ is given by
$v^c_2=1/\lambda^2$ 
and $v^c_1=(2\lambda-1)/\lambda^2$ for type--B  ($f^c>0$), and by
$v^c_2=\lambda$ and $v^c_1=1$ for type--A ($f^c=0$) transitions.
The vertical dashed lines indicate $\lambda_*$ where
\gl{e8} with $c_{n>1}\equiv0$ holds for all times; i.e.
 $r\to\infty$. The dot--dashed line
describes the linear decrease, $r\propto \lambda$, see appendix A, upon
approaching the $F_1$--model at $(v^c_1,v^c_2)=(1,0)$.     
\item[Figure 2:] Correlators $\Phi(t)$ obtained from \gls{e6}{e6b} for exponent
parameter $\lambda=0.70$. The bold line marked by a $c$ corresponds to the
bifurcation point
and the others are calculated for $v_{1,2}=v^c_{1,2}(1+\varepsilon)$,
$\varepsilon=\pm 1/4^n$, $n=0,1,\ldots$ as labelled. The time scale $t_0$ from
\gl{e10} is indicated and used in the inset to compare the curve at $v^c$ to
one crossing over to oscillations at short times (dashed line calculated for
the $F_{12}$--model from \glto{e1}{e4} with $M^{\rm  reg.}\equiv0$) and to
another with overdamped  short--time dynamics ($(s+\hat{M}^{\rm
reg.}(s))\Omega^{-2}\to \Gamma_0$, long dashes).
\item[Figure 3:] Comparison of correlators at bifurcation points of the
$F_{12}$--model for different microscopic transients and shifted
using $f^c$ are plotted versus rescaled time, with $t_0$ from
\gl{e10}.  Solutions for exponent parameters $\lambda_*=0.684$  and
$\lambda_\pm=\lambda_*\pm0.15$ as labelled are shown obtained from the
equations of structural relaxation, 
\gls{e6}{e6b} (solid  lines), and from \glto{e1}{e4}  with undamped oscillatory
($M^{\rm reg.}\equiv0$, short dashes; lines terminated for $t\lssapprox10^2
t_0$)  and with overdamped ($(s+\hat{M}^{\rm
reg.}(s))\Omega^{-2}\to \Gamma_0$, long dashes) short--time
dynamics. Dashed--dotted curves indicate the critical decay
law, $(t_0/t)^a$ with $a_+=$ 0.26 and $a_-=$ 0.39 corresponding to the
$\lambda_\pm$. For $\lambda_*$, where $a_*=1/3$, the structural relaxation
follows the critical decay for all times. Black circles (diamonds)
indicate where the structural dynamics (critical decay) curves deviate by 20 \%
in horizontal direction from the solutions with overdamped  microscopics.
\end{itemize}
\newpage\firstfigfalse
\pagestyle{empty} 
\begin{figure}[h]
  \centerline{\hspace*{0.cm}\rotate[r]{\epsfysize=18.cm 
 \epsffile{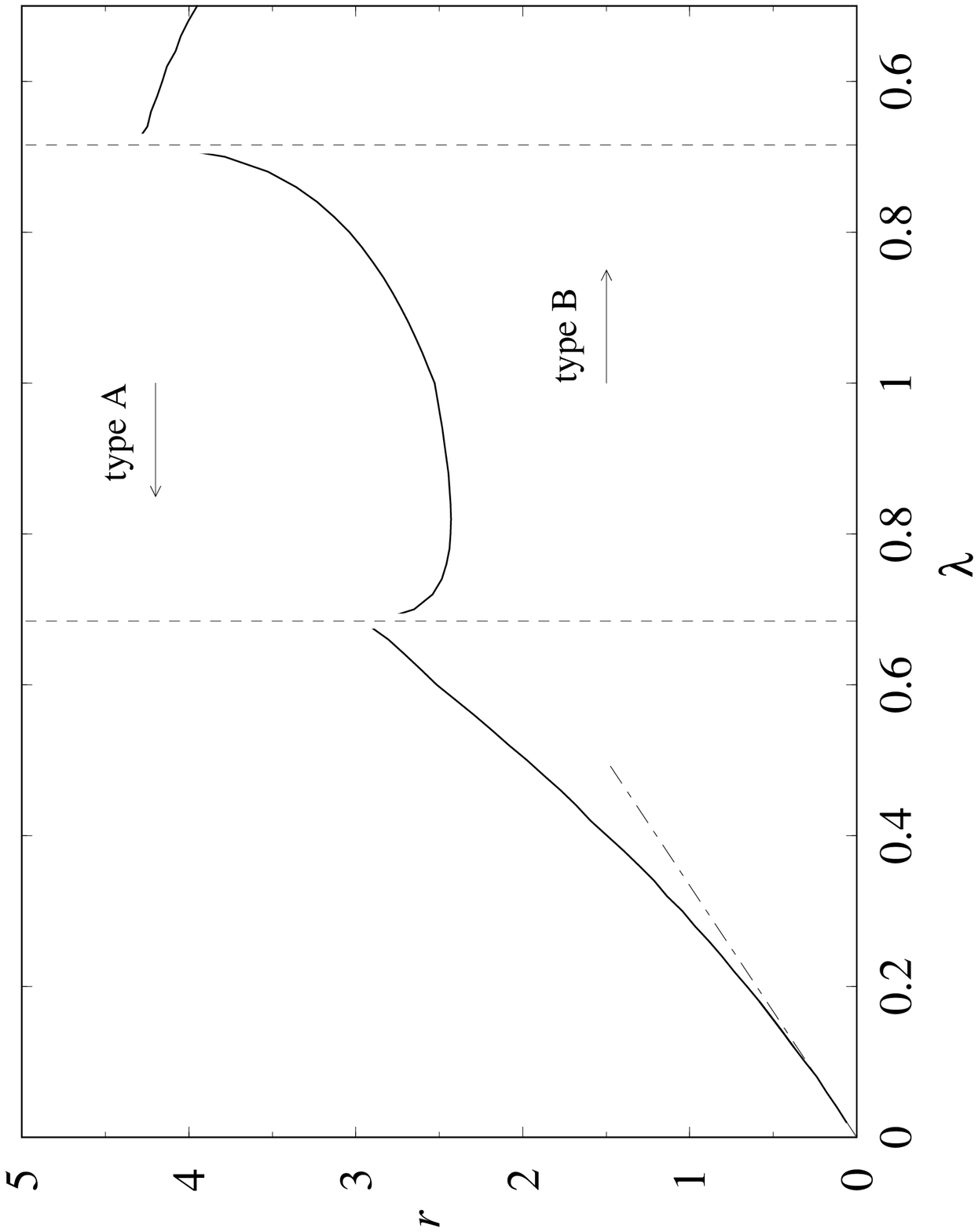}
}}
  \vspace{2cm}
  \centerline{Fuchs and Voigtmann, Equations of structural ... Fig. 1}
\end{figure}
\clearpage
\pagestyle{empty} 
\begin{figure}[h]
  \centerline{\hspace*{0.cm}\rotate[r]{\epsfysize=18.cm 
  \epsffile{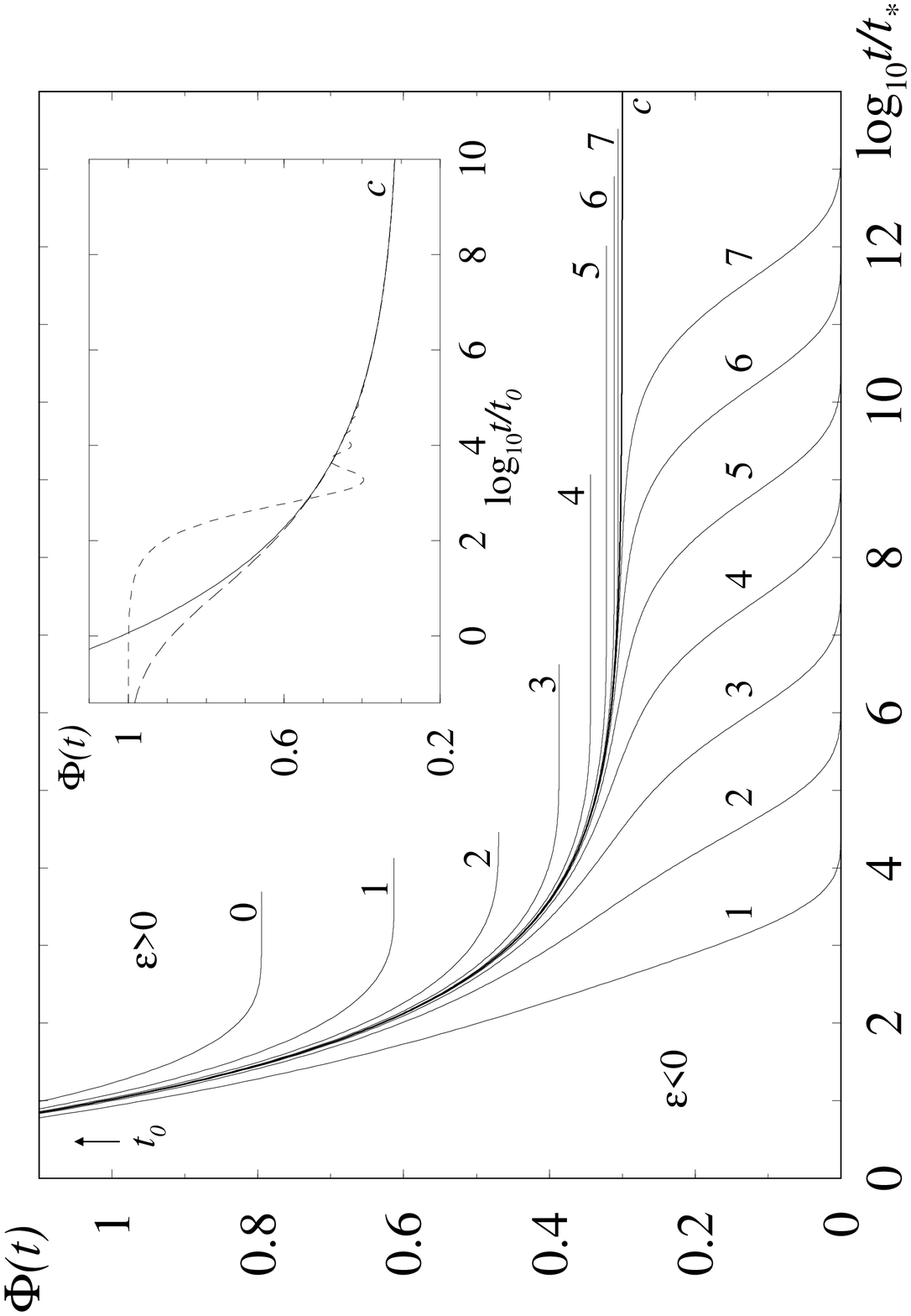}
}}
  \vspace{2cm}
  \centerline{Fuchs and Voigtmann, Equations of structural ... Fig. 2}
\end{figure}
\clearpage
\pagestyle{empty} 
\begin{figure}[h]
  \centerline{\hspace*{0.cm}\rotate[r]{\epsfysize=18.cm 
  \epsffile{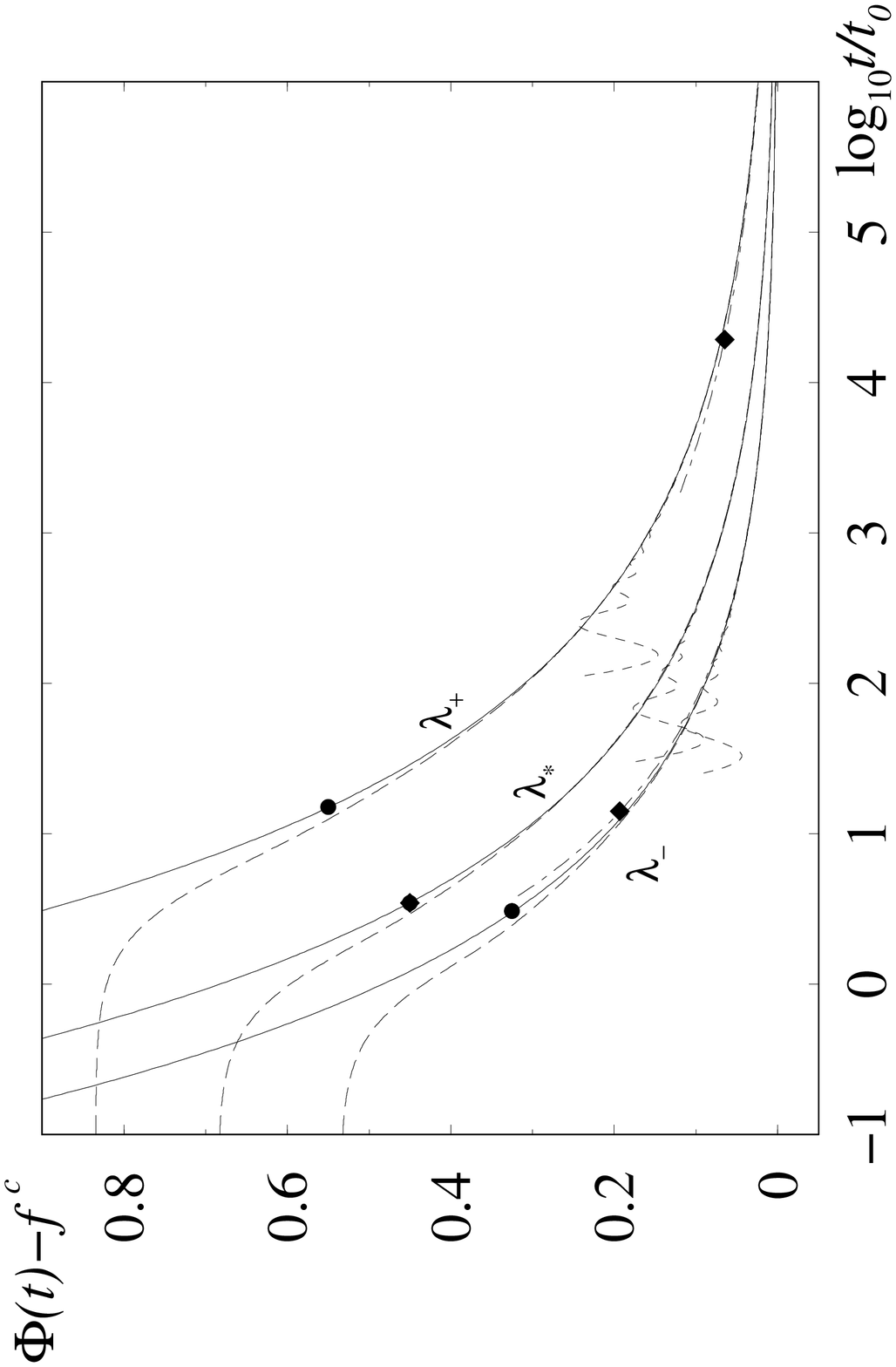}
}}
  \vspace{2cm}
  \centerline{Fuchs and Voigtmann, Equations of structural ... Fig. 3}
\end{figure}
\end{document}